\documentclass[10pt,letterpaper,twocolumn]{article} 

\usepackage{ol}
\usepackage{hyperref}
\usepackage{amsmath}
\usepackage{amssymb}

\begin{document}

\twocolumn[ 

\title{Accuracy Evaluation of an Optical Lattice Clock with Bosonic Atoms}

\author{Xavier Baillard, Mathilde Fouch\'{e}$*$, Rodolphe Le Targat, Philip G. Westergaard, Arnaud Lecallier, Yann Le Coq,
Giovanni D. Rovera, Sebastien Bize, and Pierre Lemonde}

\address{LNE-SYRTE, Observatoire de Paris, 61, avenue de l'Observatoire, 75014 Paris, France}


\begin{abstract}
We report the first accuracy evaluation of an optical lattice
clock based on the $^{1}S_{0} \rightarrow\, ^{3}P_{0}$ transition
of an alkaline earth boson, namely $^{88}$Sr atoms. This
transition has been enabled using a static coupling magnetic
field. The clock frequency is determined to be 429 228 066 418
009(32)\,Hz. The isotopic shift between $^{87}$Sr and $^{88}$Sr is
62 188 135\,Hz with fractional uncertainty $5\times10^{-7}$. We
discuss the conditions necessary to reach a clock accuracy of
$10^{-17}$ or less using this scheme.
\end{abstract}

\ocis{020.3260, 020.6580, 020.7490, 120.3940.}

 ] 

\noindent The recent advent of optical lattice clocks has opened a
very promising avenue for the future of atomic frequency
standards\cite{Takamoto05,Barber06,Letargat06,Boyd06}. Like single
ion clocks\cite{Margolis04,Dube05,Oskay06,Peik06} they allow to
efficiently cancel motional effects thanks to the lattice
confinement\cite{Lemonde05}. In addition, they can operate with a
large number of atoms and their expected ultimate performance is a
relative frequency noise well below $10^{-15}\,\tau^{-1/2}$ (with
$\tau$ the averaging time in seconds) combined with a control of
systematic effects in the $10^{-18}$
range\cite{Katopal03,Brusch06}. These new clocks use as a quantum
reference the transition between the two lowest $^1S_0$ and
$^3P_0$ states of alkaline-earth(-like) atoms: Sr, Yb, Mg, Ca, Hg,
etc. This transition is only slightly allowed by hyperfine
quenching in the fermionic isotopes of these elements, and
exhibits exquisitely narrow natural widths, in the mHz
range\cite{Courtillot03,Porsev04,Ovsiannikov06}. Most of the
experimental results demonstrated so far were obtained with
fermionic $^{87}$Sr: these include the observation of optical
resonances with Hz linewidth\cite{Boyd06}, the observation of
hyperpolarizability effects\cite{Brusch06}, and accuracy
evaluations progressively improved down to
$10^{-15}$\cite{Ludlow06,Letargat06,Takamoto06,Boyd062}.

Several proposals have been made to extend the lattice clock
scheme to bosonic isotopes. In the absence of hyperfine structure,
the true $J=0\rightarrow J=0$ transition is forbidden to all
orders for a one photon excitation, but can be enabled by adding
supplementary coupling
fields\cite{Hong05,Santra05,Taichenachev062,Zanon06,ovsiannikov07}.
The simpler structure of the clock transition in this case has
been advocated as potentially reducing sensitivity to some
systematic shifts like first order Zeeman effect or polarization
dependence of the lattice light shift. A further motivation of
such proposals is that they dramatically increase the number of
candidate species for lattice clocks experiments and that they
offer the possibility to measure isotope shifts with unprecedented
accuracy. Furthermore, interesting possibilities arise, like the
study of cold collisions in a new regime\cite{Band06}. So far,
only one of these schemes\cite{Taichenachev062}, which consists in
adding a static magnetic field, has been experimentally
demonstrated\cite{Barber06}. The experiment was performed with
$^{174}$Yb atoms and led to the observation of sub 10\,Hz
resonance linewidths and of a frequency stability below
$10^{-14}\,\tau^{-1/2}$\cite{Oates06}.

\begin{figure}[htb]
\centerline{\includegraphics[width=8.3cm]{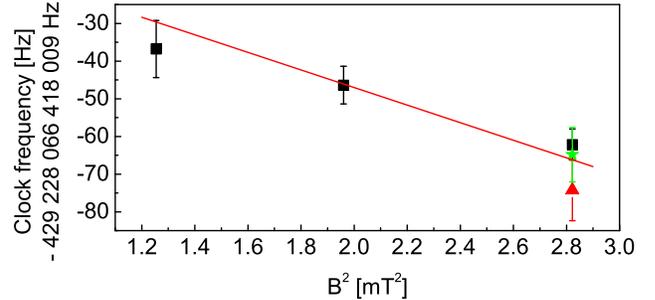}}
\caption{Dependence of the clock frequency on the static magnetic
field. The line is a fit to the experimental data with an
adjustable offset, the slope being fixed to the theoretical value
of Ref.\,\cite{Taichenachev062}. {\tiny$\blacksquare$}:
$I/I_{0}=1$, {\footnotesize$\blacktriangle$}: $I/I_{0}=0.83$,
{\scriptsize$\bigstar$}: $I/I_{0}=0.4$. } \label{fig:Zeeman}
\end{figure}

We report here the first accuracy evaluation of an optical lattice
clock with bosons, namely $^{88}$Sr. We use the same simple
magnetic coupling as in Ref.\,\cite{Taichenachev062} and report an
experimental study of the shifts induced by the coupling field.
The apparatus is derived from the $^{87}$Sr optical lattice clock
described in Ref.\,\cite{Letargat06}. Atoms are first loaded into
a magneto-optical-trap (MOT) based on the $^{1}S_{0}\rightarrow\,
^{1}P_{1}$ transition at 461\,nm, while a 1D optical lattice at
the magic wavelength and crossing the center of the MOT is
constantly on. Thanks to two lasers tuned to the lowest
$^{1}S_{0}\rightarrow\, ^{3}P_{1}$ and $^{3}P_{1}\rightarrow\,
^{3}S_{1}$ transitions, cold atoms are continuously drained into
the metastable $^{3}P_{0}$ and $^{3}P_{2}$ states. These lasers
are then switched off, and the atoms trapped in the optical
lattice are pumped back into the ground state, where they are
further cooled to $\mu$K temperatures using the narrow
$^{1}S_{0}\rightarrow\, ^{3}P_{1}$ transition at
689\,nm\,\cite{Katori99,Vogel99}. At this point, the coupling
magnetic field for the interrogation is turned on, and the clock
transition is probed using a 698 nm laser beam from an extended
cavity diode laser stabilized to a high finesse cavity. The
magnetic field is induced by two coils in Helmoltz configuration
and is parallel to the linear polarization of the probe laser
beam. The coils are fed by a power supply that switches from 0 to
6\,A in a few ms. A delay of 20\,ms is added between the end of
cooling and the beginning of the interrogation to allow for field
stabilization. The typical values used for our measurement were a
static field $B_{0}=1.68$\,mT, an interrogation time of 20\,ms,
and an intensity $I_{0}$ of the interrogation beam seen by the
atoms of about 6\,W/cm$^{2}$. The transition probability is
finally measured by detecting the populations of the $^{1}S_{0}$
and $^{3}P_{0}$ states after interrogation.

\begin{figure}[htb]
\centerline{\includegraphics[width=8.3cm]{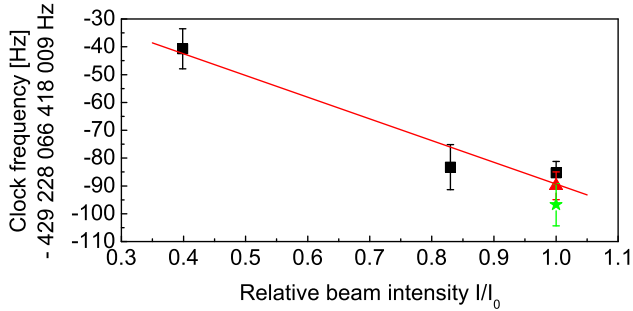}}
\caption{Light shift due to the interrogation laser for different
values of the magnetic field. {\tiny$\blacksquare$}: 1.68\,mT,
{\footnotesize$\blacktriangle$}: 1.4\,mT, {\scriptsize$\bigstar$}:
1.12\,mT. $I_{0}\simeq6$\,W/cm$^{2}$ corresponds to the clock
laser intensity at standard operating conditions. The line is a
linear fit to the data. } \label{fig:lightshift}
\end{figure}

We made an evaluation of the frequency shifts that are specific to
this clock configuration: the quadratic Zeeman shift and the light
shift due to the interrogation laser. Fig.\,\ref{fig:Zeeman} shows
the clock frequency as a function of the square of the coupling
magnetic field. The latter is calibrated to within 1\% by
measuring the linear Zeeman shift of $^{87}$Sr. Also plotted
(line) is the expected dependence on the magnetic field as
calculated in Ref.\,\cite{Taichenachev062}. Our result is in
agreement with these expectations. The quadratic Zeeman shift for
a magnetic field $B_{0}=1.68$\,mT is $\Delta_{B}=-65.8(1.3)$\,Hz.

The evaluation of the light shift due to the interrogation laser
is not as straightforward. Both the matrix elements that determine
the light shift and the absolute intensity actually seen by the
atoms are difficult to determine accurately. Instead we performed
frequency measurements for various probe laser intensities which
are referenced relative to $I_{0}$. The results are plotted in
Fig.\,\ref{fig:lightshift}. For intensity $I_{0}$, the measured
light shift is $\Delta_{L}=-74(11)$\,Hz, where a conservative
uncertainty of 15\% has been assigned to the relative intensity
evaluation.

\begin{figure}[htb]
\centerline{\includegraphics[width=8.3cm]{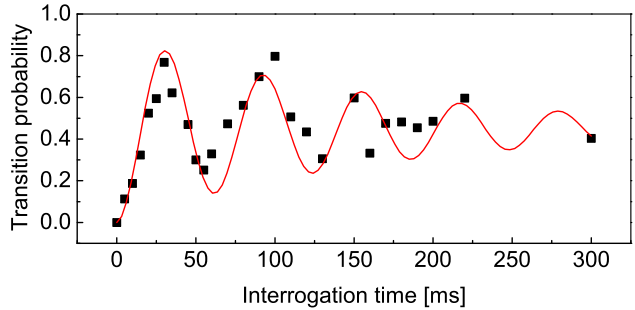}}
\caption{Experimental observation of the Rabi oscillations for
$B=1.68$\,mT at maximum intensity $I_{0}\simeq6$\,W/cm$^{2}$. The
line is a fit corresponding to classical Rabi oscillations
attenuated by an exponential decay. The frequency of the
oscillations is $16(1)$\,Hz, the time constant of the decay is
about 100\,ms which corresponds to the coherence time of the
interrogation laser.} \label{fig:Rabi}
\end{figure}

These two shifts can in principle be related to the Rabi frequency
of the transition, $\Omega/2\pi=\eta\sqrt{\Delta_{L}\Delta_{B}}$.
Ref.\,\cite{Taichenachev062} predicts $\eta=0.3$ leading to an
expected Rabi frequency $\Omega/2\pi=20.9(1.6)$\,Hz. A direct
observation of the Rabi oscillations in the same experimental
conditions is plotted on Fig.\,\ref{fig:Rabi} and gives a
frequency of $16(1)$\,Hz.

\begin{figure}[htb]
\centerline{\includegraphics[width=8.3cm]{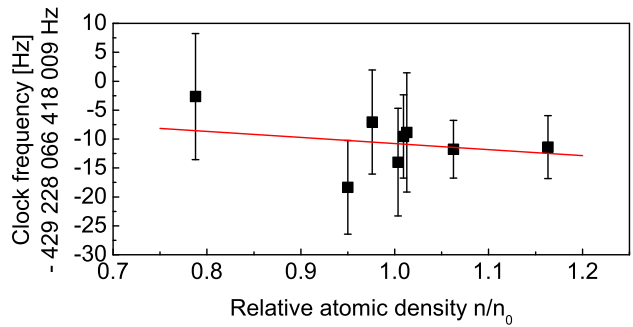}}
\caption{Clock frequency vs atomic density. All the points have
been corrected for the Zeeman and the light shifts. The line is a
linear fit to the data. We have $n_{0}=2.5 \times
10^{11}$\,at/cm$^{3}$.} \label{fig:collisions}
\end{figure}

Finally, we corrected all the data for the Zeeman and light shifts
to evaluate any possible density shift. Fig.\,\ref{fig:collisions}
shows the clock frequency as a function of the atomic density
around $n_{0}=2.5 \times 10^{11}$ at/cm$^{3}$. A linear fit to
these data gives a frequency shift compatible with zero of
$-10.4(30)$\,Hz at density $n_{0}$, or $1(3) \times
10^{-25}$\,cm$^{3}$ in fractional units. In comparison, the
density shift observed in atomic fountain clocks is of the order
of $10^{-21}$\,cm$^{3}$ for Cs \cite{Gibble93} and
$10^{-23}$\,cm$^{3}$ for $^{87}$Rb \cite{Sortais00,Fertig00}.
Another effect which has been considered is the light shift due to
the trapping field. The lattice is tuned to the magic wavelength
measured for $^{87}$Sr\cite{Brusch06}, and we made measurements
for two different depths of the trap. No detectable effect due to
the trapping light was measured.

The average value of our data corrected for systematic effects
gives a clock frequency of 429 228 066 418 008.6\,Hz with a
statistical uncertainty of 2.6\,Hz. The final uncertainty for this
measurement is 32\,Hz (see Table 1), or $7 \times 10^{-14}$ in
fractional units. This measurement in turn gives the first
accurate determination of the isotope shift for the
$^{87}$Sr-$^{88}$Sr $^{1}S_{0} \rightarrow\,^{3}P_{0}$ transition,
$\nu_{88}-\nu_{87}=$ 62 188 135.4\,Hz with relative uncertainty $5
\times 10^{-7}$.

\begin{table}
  \centering
  \caption{Uncertainty budget. All numbers are in Hz. The values are given in conditions ($I_0$, $B_0$, $n_0$).}\begin{tabular}{ccc} \\ \hline
  \hline
    Systematic effects & Correction & Uncertainty \\ \hline
    Quadratic Zeeman shift & 65.8 & 1.3 \\
    Clock laser light shift & 74.1 & 11.2 \\
    Cold collisions shift & 10.4 & 30 \\
    Blackbody radiation shift & 2.4 & $\ll$ 1 \\
    Statistical uncertainty & - & 2.6 \\ \hline
    Total & 152.7 & 32 \\ \hline \hline
  \end{tabular}
\end{table}

Our measurements validate the possibility of measuring the
$^{1}S_{0} \rightarrow \, ^{3}P_{0}$ transition for $^{88}$Sr at a
metrological level using a coupling static magnetic field. To
reach the goal of an ultimate accuracy below $10^{-17}$,
significant improvements have to be made. The first point is a
refined study of the effect of collisions between cold atoms.
Besides that, a control of the relative laser intensity to within
1\% and of the magnetic field to within a few $\mu$T is certainly
doable with our current setup. However, with our 20\,Hz Rabi
frequency, the accuracy would be still be limited to a few
$10^{-15}$, even if the collisional shift turns out not to be
problematic at that level. A $10^{-17}$ accuracy would therefore
require to work in a narrower linewidth (lower Rabi frequency)
regime. Assuming a state-of-the-art laser\cite{Young99}, we could
lower the Rabi frequency down to 0.3\,Hz while remaining
compatible with the clock laser linewidth. Under those conditions,
the goal accuracy would be within reach for a 6\,mW/cm$^{2}$ probe
intensity and 500\,$\mu$T magnetic field, but would still require
a challenging control of the probe intensity to a $10^{-3}$ level,
and of the magnetic field at better than 0.5\,$\mu$T. Magnetic
shielding as well as real time magnetic field measurement would
then probably be required. As the $^{1}S_{0}
\rightarrow\,^{3}P_{0}$ clock transition is insensitive to
magnetic field at first order, one possibility would be to use the
$^{1}S_{0} \rightarrow\,^{3}P_{1}$ transition regularly for
calibration, although great care should be taken as this
transition is sensitive to lattice trapping field intensity and
polarization effects. Alternatively, we could switch to $^{87}$Sr
from time to time.

SYRTE is Unit\'{e} Associ\'{e}e au CNRS (UMR 8630) and a member of
IFRAF. This work is supported by CNES and DGA.

\centerline{*Present address, IRSAMC, Universit\'{e} Paul
Sabatier,} \centerline{118, route de Narbonne, Toulouse, France}

\bibliographystyle{osajnl}

\end{document}